\documentclass[aps,prl,twocolumn,superscriptaddress]{revtex4}

\usepackage{graphicx}
\usepackage{dcolumn}
\usepackage{bm}

\begin{document}


\title{Torsional response and stiffening of individual multi-walled carbon nanotubes}

\author{P. A. Williams}
\affiliation{Department of Physics and Astronomy, University of 
North Carolina at Chapel Hill, Chapel Hill, NC  27599} 
\author{S. J. Papadakis}
\affiliation{Department of Physics and Astronomy, University of 
North Carolina at Chapel Hill, Chapel Hill, NC  27599} 
\author{A. M. Patel}
\affiliation{Curriculum in Applied and Materials Science, 
University of North Carolina at Chapel Hill, Chapel Hill, NC 
27599} 
\author{M. R. Falvo}
\affiliation{Curriculum in Applied and Materials Science, 
University of North Carolina at Chapel Hill, Chapel Hill, NC 
27599} 
\author{S. Washburn}
\affiliation{Department of Physics and Astronomy, University of 
North Carolina at Chapel Hill, Chapel Hill, NC  27599} 
\affiliation{Curriculum in Applied and Materials Science, 
University of North Carolina at Chapel Hill, Chapel Hill, NC 
27599} 
\author{R. Superfine}
\affiliation{Department of Physics and Astronomy, University of 
North Carolina at Chapel Hill, Chapel Hill, NC  27599} 

\date{\today}

\begin{abstract}
We report on the characterization of torsional oscillators which 
use multi-walled carbon nanotubes as the spring elements. Through 
atomic-force-microscope force-distance measurements we are able to 
apply torsional strains to the nanotubes and measure their 
torsional spring constants and effective shear moduli.  We find 
that the effective shear moduli cover a broad range, with the 
largest values near the theoretically predicted value. The data 
also suggest that the nanotubes are stiffened by repeated flexing. 
\end{abstract}


\maketitle 


The motivating vision of the future for nano-electro-mechanical 
systems (NEMS) \cite{Craighead00,Erbe01,Cleland98,Yang01} is the 
emergence of high frequency, high sensitivity devices which are 
commercially useful and also will serve as nanolabs for 
fundamental investigations into physics of the mesoscale.  The 
goal of high mechanical resonance frequency along with high 
sensitivity (high quality factor $Q$) requires that devices are  
small (low mass), stiff (high elastic modulus) and structurally 
defect free (low phonon scattering). Carbon nanotubes (CNTs) may 
provide several advantages over etched bulk semiconductors in this 
regard. They are naturally nanometer-scale objects, so little 
processing of the CNT itself is required. Furthermore, their 
surfaces are atomically ordered, and they have relatively low 
chemical reactivity, so they may not suffer from some of the 
surface-to-volume ratio issues that limit the $Q$ of 
semiconductor-based devices \cite{Lifshitz00,Carr99}. 

Because of these factors as well as their unique mechanical and 
electrical properties \cite{Postma01,Bachtold01}, CNTs may be  
ideal candidates for use in NEMS. Nevertheless, there have been 
only a handful of reports of CNTs used in an electromechanical 
setting \cite{Poncharal99,Kim99}.  We report here on the 
incorporation of CNTs into nano-electro-mechanical devices which 
allows the direct measurement of the CNTs' torsional properties. 

\begin{figure}[tb]
\centerline{ 
\includegraphics[width=2.9in, trim=0.0in 0.0in 0.0in 0.0in]{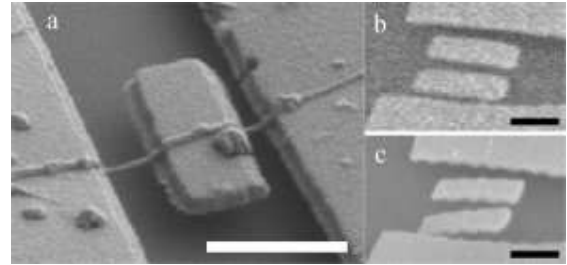}
} 
\caption{a)  Single-paddle torsional oscillator with MWNT spring-element. 
b)  A double-paddle oscillator.  c)  One end of one of the paddles 
of the device in panel b has been pushed to the substrate.  The 
deflection in the other paddle indicates that the MWNT is twisting 
uniformly along its length.  The scale bar in all three images 
corresponds to 1 $\mu$m }\vskip-0.15in 
\label{suspad}
\end{figure}

Torsional or "paddle" oscillators are commonly studied MEMS 
structures \cite{Cleland98,Evoy99,Carr00}.  We fabricate paddle 
oscillators with multi-walled carbon nanotubes (MWNTs) as the 
spring elements (Fig. \ref{suspad}).  Fabrication is described in 
detail elsewhere \cite{Williams02b}. Briefly, MWNTs are dispersed  
onto silicon wafers which have 400-500 nm of oxide.  Electron-beam 
lithography is used to pattern large metal pads over the two ends 
of each MWNT to pin them down, and a strip of metal over the 
center of each MWNT to form the paddle.  The metal is thermally 
evaporated, 15 nm of Cr followed by 100 nm of Au.  The oxide is 
etched such that the paddles are completely undercut but the 
larger pads pinning the MWNT ends are not \cite{Nygard01}.  The 
samples are then critical-point-dried. 

\begin{figure*}[tb]
\centerline{ 
\includegraphics[width=7.0in, trim=0in 0.2in 0in 0in]{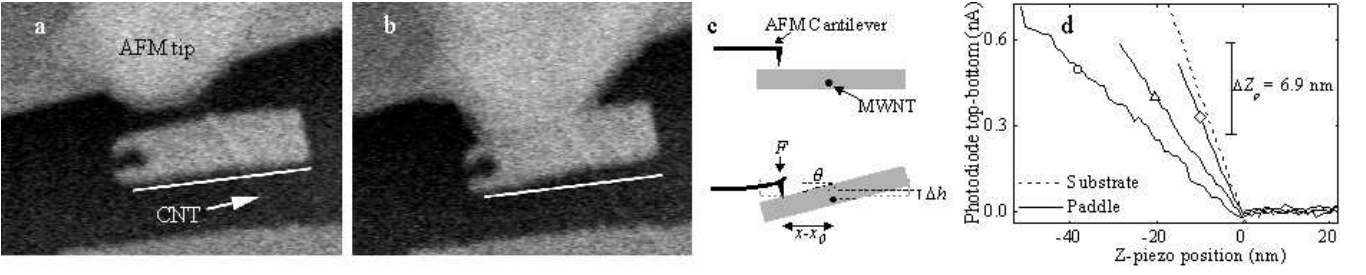} 
} 
\caption{An example of a force-distance measurement on a paddle.  a) The AFM tip
above the paddle, before the measurement is started.  b) The 
device during the measurement.  The AFM tip has deflected the left 
end of the paddle downward by 300 nm.  The right side of the 
paddle is raised, and the vertical deflection of the nanotube 
negligible.  c)  A schematic of the cantilever and paddle during 
measurement.  d)  Data from a force-distance curve on the bare 
substrate and from three consecutive traces at different positions 
on device C.  The slopes of the three curves appear, with the same 
symbols, as points in Fig. \ref{pad0404}.} \vskip-0.15in 
\label{Deflectpad}
\end{figure*}

Force measurements are performed in a hybrid atomic force 
microscope (AFM)/scanning electron microscope (SEM) system; a 
Thermomicroscopes Observer AFM mounted inside a Hitachi S-4700 
cold-cathode SEM. Using coarse-translation motors, we position the 
AFM tip above a paddle such that the entire paddle is within the 
range of the AFM scan tube. Vertical force vs. distance curves are 
taken at positions along the length of the paddle.  Figure 
\ref{Deflectpad} shows an example of a force vs. distance curve in 
progress, and typical curves resulting from such measurements. In 
all of our analyses, we use the data taken while the tip is being 
extended towards the sample. The data taken during tip retraction 
are similar. The AFM cantilevers are Si Nanosensor cantilevers 
with nominal resonance frequencies of 70-80 kHz and force 
constants $K_c =  1-3$ N/m.   

With the AFM we can apply forces to the paddles and measure their 
displacements.  We use these data to deduce the torsional 
properties of the MWNT spring-elements.  Before performing a 
quantitative analysis, we check the assumption that the MWNTs 
twist uniformly along their lengths, rather than being strained in 
a localized region.  We fabricated a device with two paddles 
suspended from one MWNT (Fig. \ref{suspad}b).  When one of the 
paddles was tilted with the AFM tip such that its end was pinned 
to the substrate, the other deflected by about half as much (Fig. 
\ref{suspad}c).  The suspended portions of the MWNT are all of 
similar length, so this suggests that the MWNT is uniformly 
strained. 

To measure force quantitatively with an AFM, we must characterize 
the AFM cantilever.  We measure its dimensions with the SEM and 
its resonance frequency with the AFM.  These values are used to 
calculate the spring constant ($K_c$) of the cantilever.  Also, 
vertical force-distance traces, which result in linear plots of 
photodiode signal vs. piezo-distance-traveled ($Z_p$), are taken 
on the substrate.  Under the assumption that the substrate is 
effectively an infinitely hard surface, the slopes of these traces 
($S_{sub}$) give us the relationship between the detector signal 
and the cantilever deflection (Fig. \ref{Deflectpad}d) 
\cite{Luthi95}.  

To characterize an oscillator, vertical force-distance curves are 
taken on the paddle itself, which yield traces with slopes 
$S_{pad}$ (Fig. \ref{Deflectpad}d).  The force applied to the 
paddle is $F = K_cZ_p{S_{pad}\over S_{sub}}$ and the vertical 
displacement of the point of contact with the AFM tip is $\Delta Z 
= Z_p(1-{S_{pad}\over S_{sub}})$.  The torsional compliance of the 
MWNT can be described by a torsional spring constant ($\kappa$), 
which relates the applied torque ($T = F(x-x_0)$, where $x-x_0$ is 
the lever arm from the axis of the MWNT) to the angular deflection 
($\theta$) of the paddle.  The vertical compliance can be 
characterized by a vertical spring constant ($K_z$), which relates 
$F$ to the vertical displacement ($\Delta h$) of the MWNT 
pivot-point \cite{paddlestiff}: 
\begin{equation}
\label{torqueeqn}
T = \kappa \theta,\;\; F = K_z\Delta h. 
\end{equation}
$\Delta h$ and $\theta$ combine to yield the displacement of the 
point of contact with the AFM tip, $\Delta Z = \Delta h + 
(x-x_0)\theta$.  Combining these equations, we obtain: 
\begin{equation}
\label{slopeseqn}
S_{pad} = \frac{S_{sub}}{1+\frac{K_c}{K_z} + 
\frac{K_c}{\kappa}(x-x_0)^2}. 
\end{equation}
On each paddle, we measure the slopes from a series of 
force-distance curves at various positions $x$.  Equation 
\ref{slopeseqn} is fit to the resulting points to yield $K_z$ and 
$\kappa$. 

\begin{figure}[tb]
\centerline{ 
\includegraphics[width=3.2in,trim= 0 1.0in 0 0]{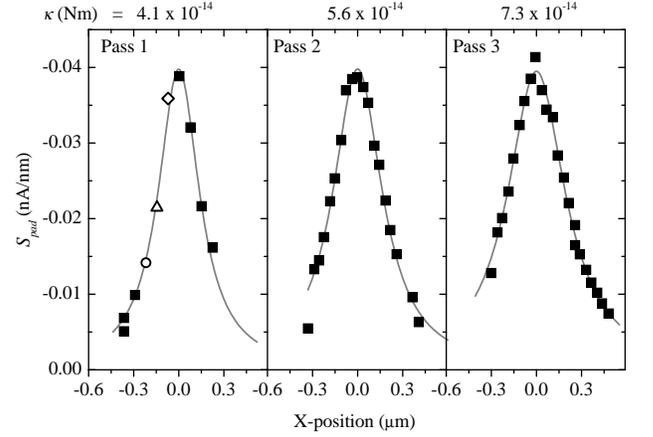}  
} 
\caption{Vertical force-distance data from paddle C, shown with fits of 
Eqn. \ref{slopeseqn}. The curves are shifted on the $X$-axis such 
that $x_0 = 0$. The panels from left to right show three 
consecutive sets of data taken on one paddle.}\vskip-0.15in 
\label{pad0404}
\end{figure}

The panels in Fig. \ref{pad0404}, from left to right, depict three 
sets of data consecutively measured on device C.  It is 
surprising, therefore, that $\kappa$ changes from one panel to the 
next. A rise in $\kappa$ is seen in all of the devices we measure, 
and the percent change in $\kappa$ from one pass to the next is 
roughly correlated with the number of deflections performed in the 
previous pass.  $\kappa$ values for some of our measurements are 
summarized in Table \ref{tab-moduli}. 

\begin{figure}[tb]
\centerline{ 
\includegraphics[width=2.4in,trim=0 0.5in 0 0]{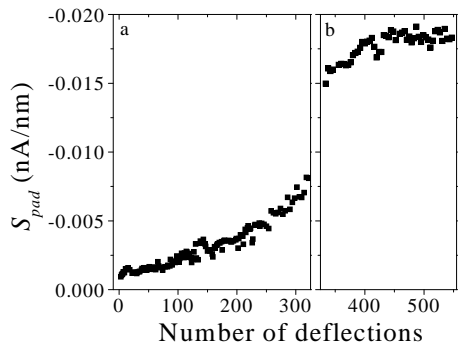}
} 
\caption{Repeated force curves taken on paddle D, showing a factor
of $\sim 10$ increase in stiffness.   (a) The AFM tip was kept in 
one place for curves 1 to 330, after which its position on the 
paddle was changed by about 50 nm, (b) where it remained for 
subsequent measurements.} \vskip-0.25in 
\label{stiffen}
\end{figure}

We also positioned the AFM tip near the end of paddle D and 
repeatedly performed vertical force-distance measurements (Fig. 
\ref{stiffen}).  After 330 force curves, the tip was moved away 
from the device and then returned to within 50 nm of the original 
position. In spite of the break in the data, the trend is clear; 
the effective torsional stiffness of the paddle increases until 
about 400 deflections are performed, and then it saturates, 
showing a net change in stiffness of about an order of magnitude.  

In principle, the MWNT could be accruing amorphous carbon 
deposited during SEM imaging of the device, but this is unlikely 
to explain its change in stiffness.  In Fig. \ref{pad0404}, we 
imaged the device during pass 1, but we turned off the electron 
beam for the  next two passes, and the stiffness still increased.  
For the data in Fig. \ref{stiffen}, the electron beam was off for 
most of the measurement, but the stiffness increases smoothly.  
After the measurements on the paddles were completed, we used the 
AFM to image the edges of the regions the SEM had been scanning, 
and saw no topographical step.  The AFM has a vertical resolution 
on the order of 1 nm, so if a significant thickness of carbon had 
been deposited during the measurement, it would have appeared as a 
topographical feature in the AFM image. To show that the 
stiffening is not an instrumental artifact, both $S_{sub}$ and the 
cantilever resonance frequency were measured before and after each 
experimental run, and they showed no significant change.  Also, we 
repeatedly deflected one AFM cantilever with another $\sim 600$ 
times and saw no change in stiffness.  This implies that the 
repeated twisting of the nanotube is making it stiffer!  The shear 
strain applied to the MWNT was typically $\sim 0.5\%$.  The 
applied tensile strain was very small.

\begin{table}[tb]
\caption{Summary of MWNT torsional spring constants ($\kappa$), outer radii ($r_{out}$),
and effective ($G_e$) and shell ($G_s$) shear moduli (See the 
text).} 
\label{tab-moduli}
\begin{tabular}{|c|c|c|c|c|}
\hline 
 Device/pass &      $\kappa$          &$r_{out}$&   $G_e$  &  $G_{s}$   \\
             &      ($10^{-14}$ Nm)   & (nm)    &   (GPa)  &   (GPa)   \\ \hline \hline                   
     A/1     &          15            &  15     &    600   &           \\ \hline
     B/1     &          2.4           &  16     &    60    &   830     \\
     B/2     &          4.5           &         &    120   &           \\
     B/3     &          4.6           &         &    120   &           \\
     B/4     &          10            &         &    280   &           \\
     B/5     &          22            &         &    590   &           \\
     B/6     &          46            &         &    1200  &           \\ \hline
     C/1     &          2.5           &  18     &    15    &   210     \\
     C/2     &          3.4           &         &    20    &           \\
     C/3     &          4.4           &         &    26    &           \\ \hline
 D\footnotemark[1]/i&     1.4           &  16     &  30    &   430   \\
 D\footnotemark[1]/f&     17            &         &  400   &         \\ \hline
\end{tabular} 
\footnotetext[1]{A full pass was not made on paddle D. $(x-x_0)$ 
was estimated from SEM images, and used in Eqn. \ref{slopeseqn} to 
calculate $\kappa$.  The $\kappa$, $G_e$, and $G_s$ given are (i) 
for the first few deflections and (f) for the saturation (See the 
text and Fig. \ref{stiffen}.)} 
\end{table}

Most existing studies focus on the tensile properties of CNTs; we 
look to these results to gain some insight. Experiments have shown 
that MWNTs can withstand tensile strains of between 10\% and 20\% 
before breaking \cite{Falvo97,Yu00}, although details of the 
structural behavior before the actual break are not known. 
Simulations of single-walled nanotubes under tension predict that 
Stone-Wales defects can form at tensile strains of $\sim 5\%$ and 
defect motion may occur at tensile strains as low as $\sim 3\%$ 
\cite{Nardelli98}. We are repeatedly applying shear strains of 
about $\sim 0.5\%$ to the MWNTs, and few times in each pass 
approaching $\sim 1\%$.  There are two structural changes that 
could stiffen the MWNTs.  The individual shells of the MWNTs could 
be stiffened or the mechanical coupling between different shells 
could be increased.  It is unlikely that changes in the connection 
between the nanotube and either the pinning metal or the paddle 
would result in an increase in device stiffness. Mechanical 
flexing of those joints would be expected to weaken them, which 
would result in a decrease in measured stiffness. 

From the $\kappa$, and the paddle dimensions, we can calculate the 
devices' expected torsional resonant frequencies, which are in the 
1-10 MHz range.  In principle, we should also be able to extract 
$K_z$ from our fits to Eqn. \ref{slopeseqn}.  For most of the 
devices, however, including device C shown in Fig. \ref{pad0404}, 
$S_{pad}$ at $(x-x_0)=0$ is within experimental error of 
$S_{sub}$.  This means that there is negligible vertical 
deflection of the paddle, and that $K_c/K_z \sim 0$.  We cannot 
accurately measure $K_z$ using a cantilever with  $K_c << K_z$. 
Nevertheless, given our experimental uncertainty in $K_c$, we can 
estimate a lower bound for $K_z$.  We return to this point below 
in our discussion of moduli. 

In order to make comparisons between MWNTs of different 
dimensions, we use the continuum mechanics model for the shear 
($G$) and Young's ($E$) moduli.  We use beam bending equations 
\cite{Young89} combined with Eqs. \ref{torqueeqn}: 
\begin{equation}
\label{KapvsGeqn}
\kappa=\frac{{\pi}(r_{out}^4-r_{in}^4)G}{2l},\;\; K_z = 
\frac{48\pi E(r_{out}^4-r_{in}^4)}{l^3}, 
\end{equation}
where $r_{out}$ and $r_{in}$ are the outer and inner radii and $l$ 
are the lengths of the suspended MWNT sections.  Since the moduli 
depend on $r_{out}^4-r_{in}^4$, taking $r_{in}$ to be 0 yields 
moduli only $\sim 6\%$ smaller than taking $r_{in}/r_{out}$ to be 
$\sim 0.5$.  Based on TEM observations, $r_{in}$ of our arc-grown 
MWNTs are typically much less than half of $r_{out}$.  We 
therefore approximate the MWNTs as solid cylinders and calculate 
{\em effective} shear moduli $G_e$, which are summarized in Table 
\ref{tab-moduli}. 

Uncertainty in $r_{out}$ contributes the bulk of the uncertainty 
in $G_e$.  AFM $r_{out}$ measurements of the metal surface under 
which the MWNTs were buried, and of MWNTs pushed down to the 
substrate with the AFM tip, varied by more than a factor of two 
for a given MWNT.  The $r_{out}$ reported in Table 
\ref{tab-moduli} were arrived at by averaging the full-width at 
half-maximum of the SEM image brightness along lines perpendicular 
to the suspended MWNTs.  Applying this technique to multiple SEM 
images of the same MWNT, taken under different imaging conditions 
(i.e. different image brightness, secondary electron detector, 
etc.), yielded a variation in the measured $r_{out}$ of about 
$20\%$, corresponding to an uncertainty in $G_e$ of about a factor 
of 2.  It is also possible that there is a systematic error in the 
estimation of $r_{out}$ from SEM images, which can exaggerate 
nanotube diameters. 

As mentioned, the MWNTs increased in stiffness as the experiment 
progressed.  Since the cause of this increase is still unknown, we 
initially discuss data from the first pass on each paddle, where 
the MWNT has been least affected by the experiment. The $G_e$ 
calculated from the first pass on each paddle range over about an 
order of magnitude.  We hypothesize that these differences are due 
to differences in the mechanical coupling between the shells of 
the MWNTs in the various devices. Cumings and Zettl have shown, by 
pulling inner shells of a MWNT out from the outer ones, that it is 
possible to have a very low intershell mechanical resistance 
\cite{Cumings00}. Yu's {\it et al.} study of the tensile failure 
modes of MWNTs shows that the individual shells are not equally 
strained in all MWNTs, since some break at widely separated places 
along the tube's length \cite{Yu00}.   Our wide range of $G_e$ 
suggests that in some MWNTs, the stress is evenly distributed 
among many shells, while in others the stress is concentrated only 
in the outer shell or few shells while the inner ones slide 
easily.  Since our MWNTs are clamped by evaporating metal onto 
their outer shells, it is conceivable that the inner shells may be 
sliding both where the MWNTs pass through the paddle and where 
they pass through the metal pinning the MWNT ends.  Observations 
have shown that some MWNTs have circular cross-sections, while 
others have polygonal cross-sections \cite{Liu94}.  Such polygonal 
MWNTs could have greater inter-shell coupling than circular MWNTs 
under rotational strain.  This would explain the large range of 
effective $G_e$ calculated from the solid rod model where the 
entire diameter of the MWNT is assumed to share the stress.  Other 
contributors to the large range of measured $G_e$ could include 
defects in the MWNTs or imperfect pinning of the nanotubes within 
the metal. 

In Ref. \onlinecite{Lu97}, using a model which assumes that all 
shells are strained, Lu predicted $G = 541$ GPa for a ten-wall 7.8 
nm-diameter CNT.  For comparison, in diamond $G = 576$ GPa and in 
graphite the basal plane $G = 440$ GPa.  Our largest measured 
$G_e$ are consistent with the theoretical value.  The experimental 
values which fall below this prediction can be explained by 
sliding of the inner shells when only the outer shell or few 
shells are clamped by the evaporated metal.  Indeed, assuming that 
only the outer shell carries the load, and using the inter-shell 
spacing as the effective thickness of that shell, we can calculate 
a shell shear modulus $G_{s}$.  Some of these results are also in 
Table \ref{tab-moduli}.  For devices B, C, and D, which initially 
have small $G_e$, $G_s$ is near the theoretical value.  The 
stiffening of the MWNTs with repeated twisting suggests that the 
inter-shell coupling can be modified.  Both devices B and D are 
stiffened by more than on order of magnitude, resulting in $G_e$  
consistent with the theoretical values.  It is noteworthy that 
device D shows a saturation at this value.  These results are 
consistent with the idea that, initially, only the outer shell is 
strained, and that repeated deflections increase the inter-shell 
coupling until all shells are strained. A more accurate $r_{out}$ 
measurement is required to confirm this conclusively.  For Young's 
modulus, if we calculate lower bounds using the lower bound values 
of the $K_z$, we find $E$ is typically greater than a few hundred 
GPa, which is consistent with previous experimental and 
theoretical results \cite{Yu00,Lu97}. 


In conclusion, we have fabricated and characterized CNT paddle 
oscillators.  We repeatedly apply torsional strains to the MWNT 
spring-elements and measure their torsional spring constants.  We 
find that the MWNTs become stiffer with repeated deflection.  The
effective shear moduli of the nanotubes calculated by 
approximating the MWNTs as solid cylinders vary by nearly two 
orders of magnitude, from near the theoretically predicted value 
to significantly less than it.  The results suggest that the 
inter-shell mechanical coupling varies widely from one MWNT to the 
next and can be modified by applying strain.

\begin{acknowledgments}
We thank the ONR and the NSF for funding for this project. 
\end{acknowledgments}


\begin{thebibliography}{10}

\bibitem{Craighead00}
H.~G. Craighead, Science {\bf 290},  1532  (2000). 

\bibitem{Erbe01}
A. Erbe, C. Weiss, W. Zwerger, and R.~H. Blick, \prl {\bf 87},  
096106  (2001). 

\bibitem{Cleland98}
A.~N. Cleland and M.~L. Roukes, Nature {\bf 392},  160  (1998). 

\bibitem{Yang01}
Y.~T. Yang {\it et~al.}, \apl {\bf 78},  162  (2001). 

\bibitem{Lifshitz00}
R. Lifshitz and M.~L. Roukes, \prb {\bf 61},  5600  (2000). 

\bibitem{Carr99}
D.~W. Carr {\it et~al.}, \apl {\bf 75},  920  (1999). 

\bibitem{Postma01}
H.~W.~C. Postma {\it et~al.}, Science {\bf 293},  76  (2001). 

\bibitem{Bachtold01}
A. Bachtold, P. Hadley, T. Nakanishi, and C. Dekker, Science {\bf 
294},  1317 
  (2001).

\bibitem{Poncharal99}
P. Poncharal, Z.~L. Wang, D. Ugarte, and W.~A. {de Heer}, Science 
{\bf 283}, 
  1513  (1999).

\bibitem{Kim99}
P. Kim and C.~M. Lieber, Science {\bf 286},  2148  (1999). 

\bibitem{Evoy99}
S. Evoy {\it et~al.}, Jour. Appl. Phys. {\bf 86},  6072  (1999). 

\bibitem{Carr00}
D.~W. Carr {\it et~al.}, \apl {\bf 77},  1545  (2000). 

\bibitem{Williams02b}
P.~A. Williams {\it et~al.}, in press. 

\bibitem{Nygard01}
J. Nyg{\aa}rd and D.~H. Cobden, \apl {\bf 79},  4216  (2001). 

\bibitem{Luthi95}
R. L\"uthi {\it et~al.}, Surf. Sci {\bf 338},  247  (1995). 

\bibitem{paddlestiff}
Bending of the metal paddle itself is negligible. The calculated 
spring 
  constant due to the bending of the paddle is one to two orders of magnitude
  larger than the effective spring constants of the CNT.

\bibitem{Yu00}
M.-F. Yu {\it et~al.}, Science {\bf 287},  637  (2000). 

\bibitem{Falvo97}
M.~R. Falvo {\it et~al.}, Nature {\bf 389},  582  (1997). 

\bibitem{Nardelli98}
M. {Buongiorno Nardelli}, B.~I. Yakobson, and J. Bernholc, \prl 
{\bf 81},  4656 
   (1998).

\bibitem{Young89}
W.~C. Young, {\em Roark's Formulas for Stress and Strain}, sixth 
ed. 
  (McGraw-Hill, New York, 1989).

\bibitem{Cumings00}
J. Cumings and A. Zettl, Science {\bf 289},  602  (2000). 

\bibitem{Liu94}
M. Liu and J.~M. Cowley, Ultramicroscopy {\bf 53},  333  (1994). 

\bibitem{Lu97}
J.~P. Lu, \prl {\bf 79},  1297  (1997). 

\end{thebibliography}

\end{document}